\documentclass[12pt]{svjour3}
\usepackage{amsmath}
\usepackage{mathptmx}
\usepackage[libertine]{newtxmath}
\usepackage[T1]{fontenc}
\usepackage{geometry}
\usepackage{color}
\usepackage{url}
\usepackage{graphicx}
\usepackage[unicode=true,pdfusetitle,
 bookmarks=true,bookmarksnumbered=false,bookmarksopen=false,
 breaklinks=false,pdfborder={0 0 1},backref=false,colorlinks=true]
 {hyperref}
\hypersetup{
 urlcolor=blue,  citecolor=blue}

\makeatletter
\usepackage[sort&compress,numbers]{natbib}

\makeatother

\begin{document}

\title{Objectivity in quantum measurement}

\author{Sheng-Wen Li \and C. Y. Cai \and X. F. Liu \and C. P. Sun}

\institute{Sheng-Wen Li \at Beijing Computational Science Research Center,
Beijing 100193, China \\
 Texas A\&M University, College Station, TX 77840 \and C. Y. Cai
\at Beijing Computational Science Research Center, Beijing 100193,
China \and X. F. Liu \at Department of Mathematics, Peking University,
Beijing 100871, China \and C. P. Sun \at Beijing Computational Science
Research Center, Beijing 100193, China\\
Graduate School of China Academy of Engineering Physics, Beijing 100193,
China }
\maketitle
\begin{abstract}
The objectivity is a basic requirement for the measurements in the
classical world, namely, different observers must reach a consensus
on their measurement results, so that they believe that the object
exists ``objectively'' since whoever measures it obtains the same
result. We find that this simple requirement of objectivity indeed
imposes an important constraint upon quantum measurements, i.e., if
two or more observers could reach a consensus on their quantum measurement
results, their measurement basis must be orthogonal vector sets. This
naturally explains why quantum measurements are based on orthogonal
vector basis, which is proposed as one of the axioms in textbooks
of quantum mechanics. The role of the macroscopicality of the observers
in an objective measurement is discussed, which supports the belief
that macroscopicality is a characteristic of classicality.
\end{abstract}

\section{Introduction}

In the classical world, the \emph{objectivity} is a basic requirement
for measurements, that is, the different observers measuring the same
object must reach a consensus on their results, so that they can be
convinced that the object exists ``objectively'' since whoever measures
it obtains the same result independent of the observers. 

But in the Copenhagen version of quantum mechanics interpretation
(QMI), this objectivity is not guaranteed since the measurement by
an observer could cause a dramatical and stochastic change in the
quantum state, namely, the ``wave-function collapse'' (WFC), and
the WFC is inevitable in the Copenhagen interpretation, because the
measuring apparatus (or observer) is treated as a purely classical
term \cite{joos_decoherence_2003}.

In the Copenhagen version of QMI, the classical apparatus is indispensable
in the constitution of quantum theory as it should be, but at the
same time it is not governed by quantum law. From the logical point
of view, this is clearly unsatisfactory \cite{gell-mann_quantum_1990,gell-mann_classical_1993,weinberg_lectures_2012,weinberg_quantum_2014,tipler_quantum_2014}.
To get rid of this inconsistent point in the quantum theory, various
``built-in'' interpretations have been proposed without postulating
the pure classicality of measuring apparatus, which leads to the WFC.
The decoherence approach \cite{zeh_interpretation_1970,zurek_pointer_1981,joos_emergence_1985,zurek_decoherence_2003,joos_decoherence_2003},
the consistent history theory \cite{griffiths_consistent_1984,griffiths_consistent_2003},
and the many-worlds interpretation (MWI) \cite{everett_relative_1957,dewitt_many-worlds_1973}
are well known representatives of these tentative solutions. Besides,
there are more drastic solutions: Bohm's hidden variable approach
\cite{bohm_suggested_1952,bohm_suggested_1952-1}, 't Hooft's deterministic
and dissipative theory \cite{t_hooft_quantummechanical_1996,t_hooft_quantum_1999,sun_algebraic_2001,liu_consequences_2001},
and Adler's trace dynamics theory \cite{adler_quantum_2004,adler_generalized_1994},
in which quantum mechanics is interpreted as an effective theory emerging
from some underlying structure.

The objectivity in quantum measurements has been discussed in the
studies of quantum Darwinism \cite{zurek_decoherence_2003,zurek_quantum_2007,zurek_quantum_2009,zurek_wave-packet_2013,zurek_quantum_2014,riedel_objective_2016}.
In the theory of quantum Darwinism, it is noticed that environments
consist of many subsystems, and observers acquire information about
a system by intercepting copies of its pointer states deposited in
fragments of the environment. In this sense, the objectivity of quantum
measurements naturally emerges. The number of copies of the data in
the environment about pointer states is the measure of objectivity.

In this paper, we ask a question: if we require a quantum measurement
be ``objective'', what constraint would be imposed by this requirement
of objectivity? 

Here we should first describe the ``objective quantum measurement''
with mathematical clarity. We understand the quantum measurement as
the establishment of the one-to-one correlation between the system
$S$ and the observer $D$, and this is encoded in the joint density
matrix $\rho_{SD}$. The objectivity requires, 1) the correlation
between the system and any observer must be the same; 2) The correlation
between any two observers should be the same as that between the observer
and the system.

These two conditions summarize the key requirement of the objectivity,
namely, all the observers could obtain the same measurement result
and verify their result with each other \cite{tipler_quantum_2014}.
With this definition, we can treat the objectivity of quantum measurement
by comparing the bipartite reductions ($\rho_{SD}$, $\rho_{SD'}$
and $\rho_{DD'}$) of the total density matrix $\rho_{SDD'}$. It
turns out that, to satisfy the above simple objectivity conditions,
any correlations obtained in the quantum measurement must be based
on orthogonal vector basis. Moreover, two observers are enough to
ensure this constraint.

It has been accepted as a basic axiom in quantum mechanics that the
states we obtain after quantum measurements are orthogonal ones. Here
our study shows this could be a natural constraint imposed by the
objectivity requirement. If some observation is not based on orthogonal
basis, its objective existence cannot be confirmed\emph{.}

In Refs.\,\cite{zurek_quantum_2007,zurek_wave-packet_2013}, it was
noticed that, by considering a faithful information transfer in the
idealistic pre-measurement, namely, $|u\rangle|A_{0}\rangle\rightarrow|u\rangle|A_{u}\rangle$,
$|v\rangle|A_{0}\rangle\rightarrow|v\rangle|A_{v}\rangle$, automatically
the unitarity of the evolution guarantees that only orthogonal basis
of the system ($\langle u|v\rangle=0$) can be well distinguished
in quantum measurements. In our study, the process how the correlations
($SD$, $SD'$ or $DD'$) are established is not concerned. By checking
whether the correlations in the total density matrix $\rho_{SDD'}$
satisfy the objectivity conditions, the orthogonality of the measurement
basis is also obtained directly. Moreover, for not only the system
$S$ but also the observers $D/D'$, the measurement basis must be
orthogonal basis. Namely, once the measurement result of a quantum
system is objectively obtained, it must have been measured in orthogonal
basis, and the measurement devices also must be working in orthogonal
basis.

\section{Quantum measurements with two or more observers}

In the QMI based on decoherence, a quantum measurement or observation
is completed in two steps:

Step 1, the pre-measurement. A non-demolition coupling of the system
$S$ to the apparatus (observer) $D$ is established and unitarily
leads to a quantum entanglement between $S$ and $D$. 

Step 2, the decoherence. The environment $E$ surrounding $S$ selects
the preferred basis $\{|\mathsf{s}_{n}\rangle\}$, and a classical
correlation is created from the quantum entanglement developed in
the pre-measurement \cite{von_neumann_mathematical_1955,zurek_pointer_1981,ollivier_quantum_2001}.

Suppose that a system $S$ initially prepared in a pure state is to
be measured. The states of $D$ and $E$ are denote by $|\mathsf{d}_{n}\rangle$
and $|\mathsf{E}_{n}\rangle$ respectively. Then in the above mentioned
Step 1 of the quantum measurement (pre-measurement), the total system
(universe) $S+D+E$ will evolve into a partially entangled state 
\begin{equation}
|\Phi_{1}\rangle=\Big[\sum_{n}c_{n}|\mathsf{s}_{n}\rangle\otimes|\mathsf{d}_{n}\rangle\Big]\otimes|\mathsf{E}\rangle,
\end{equation}
from an initial product state $|\Phi_{0}\rangle=|\psi_{S}(0)\rangle\otimes|\mathsf{d}\rangle\otimes|\mathsf{E}\rangle$.
Here, $|\mathsf{d}_{n}\rangle=U_{n}(D)|\mathsf{d}\rangle$ is a state
of $D$ correlated to the system state $|\mathsf{s}_{n}\rangle$ and
$U_{n}(D)$ is the $S$-state dependent evolution matrix. In Step
2, the environment will become entangled with the system so that the
total system reaches a GHZ type state 
\begin{equation}
|\Phi_{2}\rangle=\sum_{n}c_{n}|\mathsf{s}_{n}\rangle\otimes|\mathsf{d}_{n}\rangle\otimes|\mathsf{E}_{n}\rangle,\label{eq:Psi_t}
\end{equation}
where the environment states $|\mathsf{E}_{n}\rangle=U_{n}(E)|\mathsf{E}\rangle$
are orthogonal to one another, i.e., $\langle\mathsf{E}_{m}|\mathsf{E}_{n}\rangle=\delta_{mn}$.
By tracing over the variables of $E$, one reaches then a correlation
between $S$ and $D$ represented by the reduced density matrix $\rho_{SD}=\mathrm{tr}_{E}|\Psi\rangle\langle\Psi|$,
that is, 
\begin{equation}
\rho_{SD}=\sum_{n}|c_{n}|^{2}|\mathsf{s}_{n},\mathsf{d}_{n}\rangle\langle\mathsf{s}_{n},\mathsf{d}_{n}|,\label{eq:s-d}
\end{equation}
where $|\mathsf{s}_{n},\mathsf{d}_{n}\rangle=|\mathsf{s}_{n}\rangle\otimes|\mathsf{d}_{n}\rangle.$

The above is a sketchy description of the implementation of quantum
measurement with the help of environment $E$. It is pointed out that
one does not need to require the orthogonality among the device states
$\{|\mathsf{d}_{n}\rangle\}$ to distinguish the system states $\{|\mathsf{s}_{n}\rangle\}$.
But an ideal quantum measurement will require the orthogonality among
the device states. We will return to this topic later.

It is noticed from Eq.\,(\ref{eq:Psi_t}) that when $\{|\mathsf{d}_{n}\rangle\}$
and $\{|\mathsf{E}_{n}\rangle\}$ are both orthogonal vector sets,
from mathematical point of view, the distinction between the observer
and the environment is just nominal. Indeed, as far as the measurement
of the system state is concerned, here the state $|\Phi_{2}\rangle$
enjoys a symmetry with respect to the exchange between $|\mathsf{d}_{n}\rangle$
and $|\mathsf{E}_{n}\rangle$. Thus boundary between the observer
and the environment is not inherent in the current decoherence approach.
It has been stressed by Zurek that in the decoherence approach the
environment has been recognized as a witness of the measurement, it
essentially plays the role of another measuring device or observer,
and a large environment with redundancy of degrees of freedom can
be divided into several portions, which could be regarded as observers
\cite{ollivier_objective_2004}.

It is thus not unnatural if we replace the environment with another
observer and consider a scheme of quantum measurement with two observers.
In this scheme, the total system is made up of a system $S$ and two
observers $D$ and $D'$. These two observers can be also regarded
as two fractions in the frame of quantum Darwinism. Let $\{|\mathsf{d}_{n}\rangle\}$
and $\{|\mathsf{d}_{n}'\rangle\}$ be two bases of the state spaces
of $D$ and $D'$ respectively. The quantum measurement is then implemented
through a tripartite decomposition 
\begin{equation}
|\Psi\rangle=\sum_{n}c_{n}|\mathsf{s}_{n}\rangle\otimes|\mathsf{d}_{n}\rangle\otimes|\mathsf{d}_{n}'\rangle.
\end{equation}
In this case, both the reduced density matrices $\rho_{SD}=\mathrm{tr}_{D'}|\Psi\rangle\langle\Psi|$
and $\rho_{SD'}=\mathrm{tr}_{D}|\Psi\rangle\langle\Psi|$ characterize
a correlation between the system $S$ and an observer $D/D'$. And
$\rho_{DD'}=\mathrm{tr}_{S}|\Psi\rangle\langle\Psi|$ gives the correlation
between the two observers, which makes it possible to compare their
results. If $|\mathsf{s}_{n}\rangle$ are orthogonal states of $S$,
the correlation 
\begin{equation}
\rho_{DD'}=\mathrm{tr}_{S}|\Psi\rangle\langle\Psi|=\sum_{n}|c_{n}|^{2}|\mathsf{d}_{n},\mathsf{d}_{n}'\rangle\langle\mathsf{d}_{n},\mathsf{d}_{n}'|\label{eq:d-d'}
\end{equation}
has a classical form. If $|\mathsf{s}_{n}\rangle$ are not orthogonal
states of $S$, there will not be a perfect classical correlation
between the two observers as above. Instead, it reads
\begin{equation}
\tilde{\rho}_{DD'}=\rho_{DD'}+\sum_{m\neq n}c_{m}^{*}c_{n}|\mathsf{d}_{n},\,\mathsf{d}_{n}'\rangle\langle\mathsf{d}_{m},\,\mathsf{d}_{m}'|\cdot\langle\mathsf{s}_{m}|\mathsf{s}_{n}\rangle.
\end{equation}
Here the term $\tilde{\rho}_{DD'}-\rho_{DD'}$ will negatively influence
the comparison between the results of the two observers. This is a
hint that non-orthogonal states can not be distinguished objectively.
This point will be made clear later after a definition of measurement
related objectivity is proposed mathematically.

We observe that partially tracing is omnipresent in the domain of
quantum measurement. Physically it should imply doing some average
or coarse-graining by the Born rule. With this remark we end this
section.

\section{Objectivity of quantum measurement}

Now we see that the quantum measurement is understood as the establishment
process of the system-observer correlation, which is encoded in the
bipartite density matrices. With this consideration, we can discuss
the objectivity requirement for quantum measurements with mathematical
clarity.

As we mentioned before, the objectivity is a basic requirement for
measurements in the classical world. It at least has two basic requirements,
i.e., the different observers should obtain the same result, and they
can check their result with each other. Since the quantum measurement
is understood as the establishing process of correlations, we can
verify whether this objectivity requirement is satisfied by checking
the bipartite density matrices $\rho_{SD}$, $\rho_{SD'}$ and $\rho_{DD'}$.
These three density matrices must have the same form to guarantee
they encode the same correlation.

Therefore, the above requirements can be summarized in to the following
three \emph{objectivity conditions}: \begin{subequations} 
\begin{align}
\rho_{SD} & =\sum_{n}p_{n}|\mathsf{s}_{n},\,\mathsf{d}_{n}\rangle\langle\mathsf{s}_{n},\,\mathsf{d}_{n}|,\label{eq:rho-SD}\\
\rho_{SD'} & =\sum_{n}p_{n}|\mathsf{s}_{n},\,\mathsf{d}_{n}'\rangle\langle\mathsf{s}_{n},\,\mathsf{d}_{n}'|,\label{eq:rho-SD'}\\
\rho_{DD'} & =\sum_{n}p_{n}|\mathsf{d}_{n},\,\mathsf{d}_{n}'\rangle\langle\mathsf{d}_{n},\,\mathsf{d}_{n}'|.\label{eq:rho-DD'}
\end{align}
 \end{subequations} These three density matrices have the same form.
The first two equations mean the observers $D$ and $D'$ establish
the same correlation with the system $S$. The third equations means
$D$ and $D'$ compare their result and reach a consensus.

Notice that the correlations are established based on the basis $\{|\mathsf{s}_{n}\rangle\}$,
$\{|\mathsf{d}_{n}\rangle\}$ and $\{|\mathsf{d}_{n}'\rangle\}$,
but here we do not require them to be orthogonal vector sets. Usually
this orthogonality of measurement basis is presumed as a basis principle
in priori. Now, through the following two propositions,  we are going
to show that the orthogonality of the basis $\{|\mathsf{s}_{n}\rangle\}$,
$\{|\mathsf{d}_{n}\rangle\}$ and $\{|\mathsf{d}_{n}'\rangle\}$ is
a natural result, if we require the quantum measurement must satisfy
the above three objectivity conditions (\ref{eq:rho-SD}-\ref{eq:rho-DD'}).

\vspace{0.2cm}

\textbf{Proposition\,1} For a tripartite density matrix $\rho_{SDD'}$,
if its reduced matrices $\rho_{SD}=\mathrm{tr}_{D}[\rho_{SDD'}]$
and $\rho_{SD'}=\mathrm{tr}_{D}[\rho_{SDD'}]$ have the forms of
\begin{align}
\rho_{SD} & =\sum_{n}p_{n}|\mathsf{s}_{n},\,\mathsf{d}_{n}\rangle\langle\mathsf{s}_{n},\,\mathsf{d}_{n}|,\label{eq:rho_AB}\\
\rho_{SD'} & =\sum_{n}p_{n}|\mathsf{s}_{n},\,\mathsf{d}_{n}'\rangle\langle\mathsf{s}_{n},\,\mathsf{d}_{n}'|,\label{eq:rho_AC}
\end{align}
then there exists an orthonormal vector set $\{|\varPhi_{i}\rangle\}$,
such that the tripartite $\rho_{SDD'}$ can be written as
\begin{align}
\rho_{SDD'} & =\sum_{i}\lambda_{i}|\varPhi_{i}\rangle\langle\varPhi_{i}|,\qquad\lambda_{i}>0\nonumber \\
|\varPhi_{i}\rangle & =\sum_{n}\mathsf{C}_{n}^{(i)}|\mathsf{s}_{n},\,\mathsf{d}_{n},\,\mathsf{d}_{n}'\rangle.\label{eq:rho_ABC}
\end{align}
Here $\{|\mathsf{s}_{n}\rangle\}$, $\{|\mathsf{d}_{n}\rangle\}$
and $\{|\mathsf{d}_{n}'\rangle\}$ are complete basis sets for the
Hilbert space ${\cal H}_{S}$, ${\cal H}_{D}$ and ${\cal H}_{D'}$
respectively, but not necessarily orthogonal ones.

\vspace{0.4cm}

We leave the proof in the appendix. Any density matrix like $\rho_{SDD'}$
can be diagonalized, but it is worth noticing that this proposition
implies a strong constraint on the eigen basis $\{|\varPhi_{i}\rangle\}$,
namely, they must have a GHZ-like form {[}Eq.\,(\ref{eq:rho_ABC}){]}
(here $\{|\mathsf{s}_{n}\rangle\}$, $\{|\mathsf{d}_{n}\rangle\}$,
$\{|\mathsf{d}_{n}'\rangle\}$ may not be orthogonal basis). 

Indeed the conditions in the above Proposition\,1 can be replaced
by any two of the three objectivity conditions (\ref{eq:rho-SD}-\ref{eq:rho-DD'}),
and the conclusion is the same. If we consider some more properties
of quantum measurements, we will find that the constrained form of
$\rho_{SDD'}$ {[}Eq.\,(\ref{eq:rho_ABC}){]} imposed by Proposition\,1
can be further strengthened.

As we mentioned before, the quantum measurement is understood as the
correlation establishing process by the unitary transformation. In
the idealistic case, the initial state of the observer $D/D'$ is
prepared in a pure state. The initial state of the system $S$ to
be measured is arbitrary, namely, it can be an either pure or mixed
state. But the unitary transformation to establish the pre-measurement
should be the same for any initial state of $S$ in a specific quantum
measurement process, $\rho_{t}=U\rho_{0}U^{\dagger}$. 

Therefore, for the same pre-measurement process, pure and mixed initial
states should have equal position in the constraint imposed by the
objectivity requirement, since indeed the observers have no way to
tell the difference whether the initial state is pure or mixed in
this measurement process. Here we consider the initial state of $S$
is a pure state, the state $\rho_{SDD'}$ after pre-measurement should
also be a pure state, namely, the above Eq.\,(\ref{eq:rho_ABC})
should be written as $\rho_{SDD'}=|\varPhi\rangle\langle\varPhi|$,
and $|\varPhi\rangle=\sum_{n}c_{n}|\mathsf{s}_{n},\,\mathsf{d}_{n},\,\mathsf{d}_{n}'\rangle$. 

With this in mind, now we are going to prove the following proposition:

\vspace{0.2cm}

\textbf{Proposition\,2} We consider that the state $\rho_{SDD'}$
after pre-measurement is prepared from a pure initial state $|\psi_{S}\rangle\otimes|\mathsf{d}\rangle\otimes|\mathsf{d}'\rangle$
by a unitary transformation, in this case:

1) if the objectivity conditions (\ref{eq:rho-SD}, \ref{eq:rho-SD'})
hold, then $\{|\mathsf{d}_{n}\rangle\}$ and $\{|\mathsf{d}_{n}'\rangle\}$
must be orthonormal vector sets; 

2) if all the three objectivity conditions (\ref{eq:rho-SD}-\ref{eq:rho-DD'})
hold, then $\{|\mathsf{s}_{n}\rangle\}$ must also be an orthonormal
vector set.

\vspace{0.4cm}

\emph{Proof}: As we discussed above, since $\rho_{SDD'}$ is prepared
from a pure initial state by a unitary transformation, it also must
be a pure state. According to Proposition 1, it must have the form
of $\rho_{SDD'}=|\varPhi\rangle\langle\varPhi|$, and $|\varPhi\rangle=\sum_{n}c_{n}|\mathsf{s}_{n},\,\mathsf{d}_{n},\,\mathsf{d}_{n}'\rangle$
(this summation only encloses terms of $c_{n}\neq0$). Thus, its reduced
density matrices are 
\begin{align*}
\rho_{SD} & ={\rm tr}_{D'}|\varPhi\rangle\langle\varPhi|=\sum_{m,n}c_{m}^{*}c_{n}\langle\mathsf{d}_{m}'|\mathsf{d}_{n}'\rangle\cdot|\mathsf{s}_{n},\mathsf{d}_{n}\rangle\langle\mathsf{s}_{m},\mathsf{d}_{m}|,\\
\rho_{SD'} & ={\rm tr}_{D}|\varPhi\rangle\langle\varPhi|=\sum_{m,n}c_{m}^{*}c_{n}\langle\mathsf{d}_{m}|\mathsf{d}_{n}\rangle\cdot|\mathsf{s}_{n},\mathsf{d}_{n}'\rangle\langle\mathsf{s}_{m},\mathsf{d}_{m}'|.
\end{align*}
Comparing these expressions with the objectivity conditions (\ref{eq:rho-SD},
\ref{eq:rho-SD'}) immediately leads to the conclusion that $c_{m}^{*}c_{n}\langle\mathsf{d}_{m}'|\mathsf{d}_{n}'\rangle=c_{m}^{*}c_{n}\langle\mathsf{d}_{m}|\mathsf{d}_{n}\rangle=0$
when $m\neq n$. This implies $\langle\mathsf{d}_{m}|\mathsf{d}_{n}\rangle=\langle\mathsf{d}_{m}'|\mathsf{d}_{n}'\rangle=\delta_{mn}$,
i.e., both $\{|\mathsf{d}_{n}\rangle\}$ and $\{|\mathsf{d}_{n}'\rangle\}$
are orthonormal vector sets, thanks to the fact that $c_{n}$ are
nonzero complex numbers. The first part of the proposition is thus
proved, and the proof for the second part follows the same reason.
$\hfill\blacksquare$

\vspace{0.2cm}

From Proposition 2 we see that if we require the two observers $D/D'$
could obtain the same measurement result, namely, they establish the
same correlation with the system $S$ {[}objectivity conditions (\ref{eq:rho-SD},
\ref{eq:rho-SD'}){]}, their measurement basis $\{|\mathsf{d}_{n}\rangle\}$
and $\{|\mathsf{d}_{n}'\rangle\}$ must be orthonormal vector sets.
Further, if the two observers $D/D'$ could verify that they obtain
the same result by checking their own correlation $\rho_{DD'}$ {[}objectivity
condition (\ref{eq:rho-DD'}){]}, then the basis $\{|\mathsf{s}_{n}\rangle\}$
of the system $S$, which is what they measured, also must be an orthonormal
set.

Therefore, all the basis $\{|\mathsf{s}_{n}\rangle\}$, $\{|\mathsf{d}_{n}\rangle\}$
and $\{|\mathsf{d}_{n}'\rangle\}$ in the quantum measurement are
orthonormal set, and the state $|\varPhi\rangle$ is strictly a GHZ
state. It is worth noticing that this is a natural constraint imposed
by the requirement of objectivity, and we no more need to presume
in priori as a basic principle that the measurement basis must be
orthogonal sets. Once the measurement result of a quantum system is
objectively obtained, it must have been measured in orthogonal basis,
and the measurement devices also must be working in orthogonal basis.
Otherwise, the objectivity of the quantum system cannot be confirmed,
namely, non-orthogonal basis cannot be objectively measured.

It should be clear that all the results in this section can be generalized
to multi-observer cases without difficulty. We would rather not go
into the details.

\section{Ideal measurement from macroscopicality: central spin model}

To achieve the above objective measurement, we need a unitary evolution
satisfying $U\big(|\mathsf{s}_{n}\rangle\otimes|\mathsf{d},\mathsf{d}'\rangle\big)=|\mathsf{s}_{n}\rangle\otimes|\mathsf{d}_{n},\mathsf{d}'_{n}\rangle$.
This can be completed by a Hamiltonian of the non-demolition type:
$[\hat{H}_{S},\,\hat{H}_{SD}]=0,\,[\hat{H}_{S},\,\hat{H}_{SD'}]=0$
and $[\hat{H}_{SD},\,\hat{H}_{SD'}]=0$, where $\hat{H}_{SD}$ and
$\hat{H}_{SD'}$ are the interaction between $S$ and $D/D'$. And
such a unitary transformation could be achieved by a dedicate control
of the interaction time.

Besides, there is another more natural way to realize this unitary
transformation by noticing that each macroscopic observer is usually
composed of infinitely many degrees of freedom, and the orthogonality
$\langle\mathsf{d}_{m}|\mathsf{d}_{n}\rangle=\langle\mathsf{d}_{m}'|\mathsf{d}_{n}'\rangle=\delta_{mn}$
can be achieved asymptotically in the thermodynamical limit.

Consider a composite system $S+D^{(1)}+D^{(2)}+\cdots+D^{(N)}$, where
$S$ is meant to be a quantum system to be measured and $D^{(1)},D^{(2)},\cdots,D^{(N)}$
stand for ``elementary'' observers. Choose a non-demolition type
Hamiltonian $\hat{{\cal H}}$, such that an correlated state is prepared
as
\begin{equation}
|\Psi\rangle=\sum_{n}c_{n}|\mathsf{s}_{n}\rangle\bigotimes_{i=1}^{N}|d_{n}^{(i)}\rangle.
\end{equation}
Generally speaking, one cannot expect $\{|d_{n}^{(j)}\rangle\}$ to
be orthogonal vector sets without a dedicate control of the interaction
and the evolution time, and thus it is not easy to satisfy the above
objectivity requirement. But if we divide the $N$ ``elementary''
observers into two parts, i.e., by defining $|\mathsf{D}_{n}\rangle:=\bigotimes_{i=1}^{M}|d_{n}^{(i)}\rangle$
and $|\mathsf{D}_{n}'\rangle:=\bigotimes_{i=M+1}^{N}|d_{n}^{(i)}\rangle$,
then in the macroscopical limit $M\rightarrow\infty$ and $N\rightarrow\infty$
we could have 
\begin{eqnarray}
\langle\mathsf{D}_{m}|\mathsf{D}_{n}\rangle & = & \prod_{i=1}^{M}\langle d_{m}^{(i)}|d_{n}^{(i)}\rangle\rightarrow0,\nonumber \\
\langle\mathsf{D}_{m}'|\mathsf{D}_{n}'\rangle & = & \prod_{i=M+1}^{N}\langle d_{m}^{(i)}|d_{n}^{(i)}\rangle\rightarrow0,
\end{eqnarray}
for $m\ne n$ if only $|\langle d_{m}^{(i)}|d_{n}^{(i)}\rangle|<1$
for $m\ne n$, which are easy to satisfy. This means that the ``elementary''
observers $\{D^{(1)},D^{(2)},\cdots,D^{(M)}\}$ and $\{D^{(M+1)},D^{(M+2)},\cdots,D^{(N)}\}$
can be coarse-grained into two macroscopic observers $D$ and $D'$
effectively. This observation convinces us that macroscopicality may
well be regarded as a characteristic of a quantum observer.

To illustrate the above argument, let us present a concrete example
\cite{zurek_pointer_1981,sun_quantum_1993,quan_decay_2006}. In this
example, the quantum system $S$ to be measured is a central spin
with two states $|\mathsf{e}\rangle$ and $|\mathsf{g}\rangle$, and
the central spin is surrounded by another $N$ spin-$\frac{1}{2}$
particles, which serve as the above mentioned ``elementary'' observers
$D^{(1)}+D^{(2)}+\cdots+D^{(N)}$. The Hamiltonian of the total system
$S+D^{(1)}+D^{(2)}+\cdots+D^{(N)}$ reads 
\begin{equation}
\hat{{\cal H}}=\mathcal{E}|\mathsf{e}\rangle\langle\mathsf{e}|+\sum_{i=1}^{N}(\omega_{i}\hat{\sigma}_{i}^{z}+g_{i}\hat{\sigma}_{i}^{x})+|\mathsf{e}\rangle\langle\mathsf{e}|\cdot[\sum_{i=1}^{N}\eta_{i}\hat{\sigma}_{i}^{z}],
\end{equation}
where $\hat{\sigma}_{i}^{z}=|\uparrow\rangle_{i}\langle\uparrow|-|\downarrow\rangle_{i}\langle\downarrow|$
and $\hat{\sigma}_{i}^{x}=|\uparrow\rangle_{i}\langle\downarrow|+|\downarrow\rangle_{i}\langle\uparrow|$
are the Pauli matrices for the $i$-th spin.

\begin{figure}
\begin{centering}
\includegraphics[width=0.6\columnwidth]{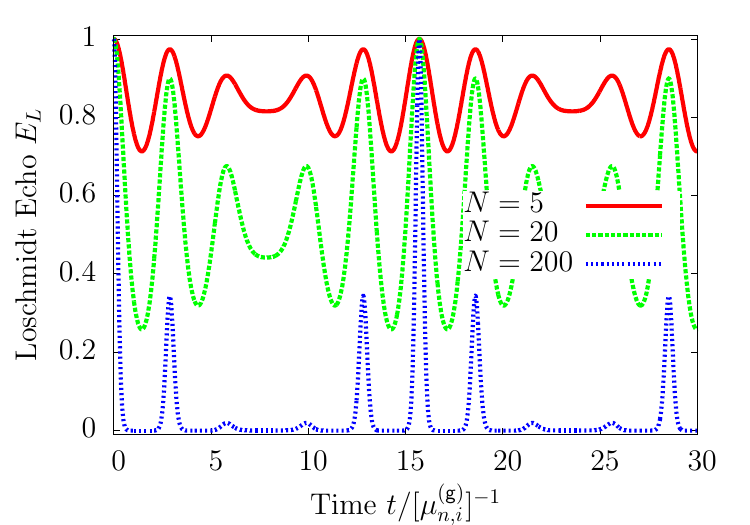}
\par\end{centering}
\caption{When the spin number $N_{1,2}\rightarrow\infty$, the Loschmidt echo
$E_{L}^{1,2}$ {[}Eq.\,(\ref{eq:E_L}){]} approaches zero. We set
$\mu_{n,i}^{(\mathsf{g})}=1$ as the energy unit, and $\mu_{n,i}^{(\mathsf{e})}=1.2$,
$g_{n,i}=0.2$.}
\label{fig-deco}
\end{figure}

In the spirit of the preceding discussion, we coarse-grain the $N$
``elementary'' observers into two macroscopic observers $D$ and
$D'$ which contains $N_{1}$ and $N_{2}$ spins respectively ($N_{1}+N_{2}=N$),
namely, we take $D=D^{(1)}+D^{(2)}+\cdots+D^{(N_{1})}$ and $D'=D^{(N_{1}+1)}+D^{(N_{1}+2)}+\cdots+D^{(N_{1}+N_{2})}$.
For clarity, we rewrite the Hamiltonians of the systems $D$ and $D'$
as 
\begin{eqnarray}
\hat{H}_{D} & = & \sum_{i=1}^{N_{1}}(\omega_{1,i}\hat{\sigma}_{1,i}^{z}+g_{1,i}\hat{\sigma}_{1,i}^{x}),\nonumber \\
\hat{H}_{D'} & = & \sum_{i=1}^{N_{2}}(\omega_{2,j}\hat{\sigma}_{2,j}^{z}+g_{2,j}\hat{\sigma}_{2,j}^{x}).
\end{eqnarray}
 It is then a routine work to check that 
\begin{eqnarray}
e^{-i\hat{{\cal H}}T}(|\mathsf{g}\rangle\bigotimes_{i=1}^{N}|\uparrow\rangle)= &  & |\mathsf{g}\rangle\otimes|\mathsf{D}_{\mathsf{g}}\rangle\otimes|\mathsf{D}_{\mathsf{g}}'\rangle,\nonumber \\
e^{-i\hat{{\cal H}}T}(|\mathsf{e}\rangle\bigotimes_{i=1}^{N}|\uparrow\rangle)= & e^{-i\mathcal{E}T} & |\mathsf{e}\rangle\otimes|\mathsf{D}_{\mathsf{e}}\rangle\otimes|\mathsf{D}_{\mathsf{e}}'\rangle,
\end{eqnarray}
where 
\begin{alignat}{2}
|\mathsf{D}_{\mathsf{g}}\rangle & =\bigotimes_{i=1}^{N_{1}}R_{1,i}^{(\mathsf{g})}(T)|\uparrow\rangle, & \quad & |\mathsf{D}_{\mathsf{g}}'\rangle=\bigotimes_{i=1}^{N_{2}}R_{2,i}^{(\mathsf{g})}(T)|\uparrow\rangle,\nonumber \\
|\mathsf{D}_{\mathsf{e}}\rangle & =\bigotimes_{i=1}^{N_{1}}R_{1,i}^{(\mathsf{e})}(T)|\uparrow\rangle, &  & |\mathsf{D}_{\mathsf{e}}'\rangle=\bigotimes_{i=1}^{N_{2}}R_{2,i}^{(\mathsf{e})}(T)|\uparrow\rangle,
\end{alignat}
with $R_{n,i}^{(\alpha)}(T)=\exp[-iH_{n,i}^{(\alpha)}T]$ for $\alpha=\mathsf{g},\,\mathsf{e}$
and $n=1,2$. Here $H_{n,i}^{(\alpha)}$ are single effective Hamiltonians
defined as follows: 
\begin{eqnarray}
H_{n,i}^{(\mathsf{g})} & = & \omega_{n,i}\hat{\sigma}_{n,i}^{z}+g_{n,i}\hat{\sigma}_{n,i}^{x},\nonumber \\
H_{n,i}^{(\mathsf{e})} & = & (\omega_{n,i}+\eta_{n,i})\hat{\sigma}_{n,i}^{z}+g_{n,i}\hat{\sigma}_{n,i}^{x}.
\end{eqnarray}
By straightforward calculation, we obtain 
\begin{align}
 & \left|\langle\mathsf{D}_{\mathsf{g}}|\mathsf{D}_{\mathsf{e}}\rangle\right|=\prod_{i=1}^{N_{1}}\langle\uparrow|[R_{1,i}^{(\mathsf{g})}(T)]^{\dagger}\cdot R_{1,i}^{(\mathsf{e})}(T)|\uparrow\rangle\label{eq:E_L}\\
= & \prod_{i=1}^{N_{1}}(1-\sin^{2}\mu_{1,i}^{(\mathsf{e})}T\cdot\sin^{2}\phi_{1,i}^{(\mathsf{e})})(1-\sin^{2}\mu_{1,i}^{(\mathsf{g})}T\cdot\sin^{2}\phi_{1,i}^{(\mathsf{g})}),\nonumber \\
 & \left|\langle\mathsf{D}_{\mathsf{g}}'|\mathsf{D}_{\mathsf{e}}'\rangle\right|=\prod_{i=1}^{N_{2}}\langle\uparrow|[R_{2,i}^{(\mathsf{g})}(T)]^{\dagger}\cdot R_{2,i}^{(\mathsf{e})}(T)|\uparrow\rangle\nonumber \\
= & \prod_{i=1}^{N_{2}}(1-\sin^{2}\mu_{2,i}^{(\mathsf{e})}T\cdot\sin^{2}\phi_{2,i}^{(\mathsf{e})})(1-\sin^{2}\mu_{2,i}^{(\mathsf{g})}T\cdot\sin^{2}\phi_{2,i}^{(\mathsf{g})}),\nonumber 
\end{align}
where 
\begin{alignat}{2}
\mu_{n,i}^{(\mathsf{e})} & =[(\omega_{n,i}+\eta_{n,i})^{2}+g_{n,i}^{2}]^{\frac{1}{2}}, & \quad & \sin\phi_{n,i}^{(\mathsf{e})}=\frac{g_{n,i}}{\mu_{n,i}^{(\mathsf{e})}},\nonumber \\
\mu_{n,i}^{(\mathsf{g})} & =[\omega_{n,i}^{2}+g_{n,i}^{2}]^{\frac{1}{2}}, & \quad & \sin\phi_{n,i}^{(\mathsf{g})}=\frac{g_{n,i}}{\mu_{n,i}^{(\mathsf{g})}},
\end{alignat}
for $n=1,2$. It is noticed that $\left|\langle\mathsf{D}_{\mathsf{g}}|\mathsf{D}_{\mathsf{e}}\rangle\right|$
and $\left|\langle\mathsf{D}_{\mathsf{g}}'|\mathsf{D}_{\mathsf{e}}'\rangle\right|$
are none other than the so called Loschmidt echoes. Let us denote
them by $E_{L}^{1}$ and $E_{L}^{2}$ respectively. From the expressions
of the Loschmidt echoes it should be clear that each product factor
is a non-negative number and smaller than $1$, thus in the thermodynamic
limit $N_{1,2}\rightarrow\infty$, for a generic $T$, we have $\left|\langle\mathsf{D}_{\mathsf{g}}|\mathsf{D}_{\mathsf{e}}\rangle\right|\simeq0$
and $\left|\langle\mathsf{D}_{\mathsf{g}}'|\mathsf{D}_{\mathsf{e}}'\rangle\right|\simeq0$
(see Fig.\,\ref{fig-deco}).

\section{Conclusions }

In this paper, we show that the requirement of objectivity indeed
could impose an important constraint on quantum measurements, namely,
if we require the quantum measurement to be objective, then the measurement
basis must be orthogonal vector sets. Usually this is presumed as
a basic principle in priori, but here we show that this can be a natural
constraint imposed by the requirement of objectivity.

The quantum measurement is understood as the establishing process
of correlations. And the objectivity requires that different observers
could obtain the same result, and they can verify with each other.
This is a very natural requirement in our classical world. Our result
implies if the quantum measurement is not based on orthogonal basis,
its objective existence cannot be confirmed, in another word, non-orthogonal
basis cannot be objectively measured.

The emergence of classicality in quantum measurement is closely related
to the objectivity condition. This point is illustrated with the central
spin model, where the $N$ ``elementary'' observers are coarse-grained
into two macroscopic observers enjoying orthogonal pointer state sets
for an ideal measurement. In this example, it is clearly seen how
classical correlations result from the macroscopical observers and
a support is provided for the belief that macroscopicality is a characteristic
of classicality.

\emph{Acknowledgement} -- This work is supported by National Basic
Research Program of China (Grant No. 2016YFA0301201 \& No. 2014CB921403),
NSFC (Grant No. 11534002) and NSAF (Grant No. U1730449 \& No. U1530401).
We thank S. M. Fei (Capital Normal University) and D. L. Zhou (Institute
of Physics, CAS) for helpful discussions. CPS also acknowledges Prof.
J\"urgen Jost for his kind invitation to visit Max Planck Institute
for Mathematics in the Sciences, where the manuscript was finally
accomplished.

\appendix

\section{Proof for the proposition 1}

\textbf{Proposition\,1} For a tripartite density matrix $\rho_{ABC}$,
if its reduced matrices $\rho_{AB}=\mathrm{tr}_{C}[\rho_{ABC}]$ and
$\rho_{AC}=\mathrm{tr}_{B}[\rho_{ABC}]$ have the forms of
\begin{align}
\rho_{AB} & =\sum_{n}p_{n}|\mathsf{a}_{n},\,\mathsf{b}_{n}\rangle\langle\mathsf{a}_{n},\,\mathsf{b}_{n}|,\label{eq:rho_AB-1}\\
\rho_{AC} & =\sum_{n}p_{n}|\mathsf{a}_{n},\,\mathsf{c}_{n}\rangle\langle\mathsf{a}_{n},\,\mathsf{c}_{n}|,\label{eq:rho_AC-1}
\end{align}
then there exists an orthonormal vector set $\{|\varPhi_{i}\rangle\}$,
such that the tripartite $\rho_{ABC}$ can be written as
\begin{align}
\rho_{ABC} & =\sum_{i}\lambda_{i}|\varPhi_{i}\rangle\langle\varPhi_{i}|,\qquad\lambda_{i}\ge0\nonumber \\
|\varPhi_{i}\rangle & =\sum_{n}\mathsf{C}_{n}^{(i)}|\mathsf{a}_{n},\,\mathsf{b}_{n},\,\mathsf{c}_{n}\rangle.
\end{align}
Here $\{|\mathsf{a}_{n}\rangle\}$, $\{|\mathsf{b}_{n}\rangle\}$
and $\{|\mathsf{c}_{n}\rangle\}$ are complete basis sets for the
Hilbert space ${\cal H}_{A}$, ${\cal H}_{B}$ and ${\cal H}_{C}$
respectively, but not necessarily orthogonal ones.

\vspace{0.2cm}

For clarity, we use $A,\,B,\,C$ here to replace the $S,\,D,\,D'$
in the main text. To prove this proposition, we need the following
lemma:

\vspace{0.2cm}

\textbf{Lemma} Let $\mathbf{P}$ be a positive definite matrix and
$\mathbf{C}$ a semi-positive one. If ${\rm tr}[\mathbf{C}\cdot\mathbf{P}]=0$,
then $\mathbf{C}$ is a zero matrix.

\emph{Proof}: We decompose the positive matrix $\mathbf{P}$ in its
eigen basis as $\mathbf{P}=\sum_{n}\lambda_{n}|n\rangle\langle n|$,
where all $\lambda_{n}>0$. Then we have $\mathrm{tr}[\mathbf{C}\cdot\mathbf{P}]=\sum_{n}\lambda_{n}\langle n|\mathbf{C}|n\rangle=0$.
To make sure $\langle n|\mathbf{C}|n\rangle=0$ for all the basis
$\{|n\rangle\}$, $\mathbf{C}$ must be a zero matrix. $\hfill\blacksquare$

\vspace{0.2cm}

With the help of the above lemma, the proof of Proposition\,1 lies
as follows.

\emph{Proof}: For the tripartite density matrix $\rho_{ABC}$, we
can always write it as the eigen spectrum decomposition $\rho_{ABC}=\sum_{i}\lambda_{i}|\varPhi_{i}\rangle\langle\varPhi_{i}|$,
where $|\varPhi_{i}\rangle$ are orthonormal basis, and $\lambda_{i}>0$
are the non-zero eigenvalues respectively. But now we could only write
down $|\varPhi_{i}\rangle$ in a general form
\begin{equation}
|\varPhi_{i}\rangle=\sum_{m=1}^{M}\sum_{n=1}^{N}\sum_{l=1}^{L}\mathsf{C}_{mnl}^{(i)}|\mathsf{a}_{m},\,\mathsf{b}_{n},\,\mathsf{c}_{l}\rangle,
\end{equation}
where $\mathsf{C}_{nml}^{(i)}$ are complex numbers. It then follows
that 
\begin{gather}
\rho_{ABC}=\sum_{{mnl\atop m'n'l'}}\varrho_{mnl,\,m'n'l'}|\mathsf{a}_{m},\mathsf{b}_{n},\mathsf{c}_{l}\rangle\langle\mathsf{a}_{m'},\mathsf{b}_{n'},\mathsf{c}_{l'}|,\nonumber \\
\varrho_{mnl,\,m'n'l'}:=\sum_{i}\lambda_{i}\cdot\mathsf{C}_{mnl}^{(i)}\overline{\mathsf{C}}_{m'n'l'}^{(i)},
\end{gather}
and the reduced density matrix $\rho_{AB}$ becomes
\begin{equation}
\rho_{AB}={\rm Tr}_{C}[\rho_{ABC}]=\sum_{{m,n\atop m',n'}}\Big(\sum_{l,l'}\varrho_{mnl,\,m'n'l'}\langle\mathsf{c}_{l'}|\mathsf{c}_{l}\rangle\Big)|\mathsf{a}_{m},\mathsf{b}_{n}\rangle\langle\mathsf{a}_{m'},\mathsf{b}_{n'}|.
\end{equation}
Comparing this with the required form of $\rho_{AB}$ {[}Eq.\,(\ref{eq:rho_AB-1}){]},
we come to the following equation 
\begin{equation}
\sum_{l,l'}\varrho_{mnl,\,m'n'l'}\langle\mathsf{c}_{l'}|\mathsf{c}_{l}\rangle=\delta_{mm'}\delta_{nn'}\cdot\delta_{mn}p_{n}.\label{eq:sum}
\end{equation}

Now we introduce two $L\times L$ matrices $\mathbf{C}^{(mn;m'n')}$
and $\mathbf{P}$, which are defined by 
\begin{equation}
[\mathbf{C}^{(mn;m'n')}]_{l,l'}=\varrho_{mnl,\,m'n'l'},\quad\mathbf{P}_{l',l}=\langle\mathsf{c}_{l'}|\mathsf{c}_{l}\rangle.
\end{equation}
With their help, Eq.\,(\ref{eq:sum}) can be written in a compact
form
\begin{equation}
\mathrm{tr}[\mathbf{C}^{(mn;m'n')}\cdot\mathbf{P}]=\delta_{mm'}\delta_{nn'}\cdot\delta_{mn}p_{n}.
\end{equation}

One notices that when $m=m'$, $n=n'$, $m\neq n$, we have
\begin{equation}
\mathrm{tr}[\mathbf{C}^{(mn;mn)}\cdot\mathbf{P}]=0.
\end{equation}
 It is easy to verify that $\mathbf{C}^{(mn;mn)}$ is a semi-positive
matrix \footnote{Obviously, the coefficient matrix $\varrho_{mnl,\,m'n'l'}$ of the
density operator $\rho_{ABC}$ is semi-positive. Notice that $[\varrho_{mnl,\,m'n'l'}]$
can be regarded as a block matrix, and $[\mathbf{C}^{(mn;mn)}]_{l,l'}=\varrho_{mnl,\,mnl'}$
is one of its principal blocks, thus $\mathbf{C}^{(mn;mn)}$ is semi-positive.}, and $\mathbf{P}$ is positive definite \footnote{For any non-zero vector $\mathbf{v}:=(v_{1},v_{2},\dots,v_{L})^{T}$,
we have $\mathbf{v}^{\dagger}\cdot\mathbf{P}\cdot\mathbf{v}=\sum_{l,l'}v_{l'}^{*}\protect\langle \mathsf{c}_{l'}| \mathsf{c}_{l}\protect\rangle v_{l}=\protect\langle\tilde{\psi}|\tilde{\psi}\protect\rangle>0$
, where $|\tilde{\psi}\protect\rangle:=\sum_{l}v_{l}| \mathsf{c}_{l}\protect\rangle$}. Therefore, according to above lemma, we know that $\mathbf{C}^{(mn;mn)}$
is a zero matrix when $m\ne n$. Thus we obtain
\begin{equation}
[\mathbf{C}^{(mn;mn)}]_{l,l}=\varrho_{mnl,\,mnl}=\sum_{i}\lambda_{i}\cdot|\mathsf{C}_{mnl}^{(i)}|^{2}=0.
\end{equation}
 Since all the $\lambda_{i}>0$ in the above summation, that leads
to 
\begin{equation}
\mathsf{C}_{mnl}^{(i)}=0,\quad\forall\,i,\,l,\,m\ne n.
\end{equation}
In the same way, by comparing with $\rho_{AC}$ {[}Eq.\,(\ref{eq:rho_AC-1}){]},
we can prove 
\begin{equation}
\mathsf{C}_{mnl}^{(i)}=0,\quad\forall\,i,\,n,\,m\ne l.
\end{equation}
Therefore, the only possible non-zero coefficients $\mathsf{C}_{mnl}^{(i)}$
are those satisfying $m=n=l$, thus, we write the coefficients as
$\mathsf{C}_{mnl}^{(i)}=\delta_{mn}\delta_{ml}\cdot\mathsf{C}_{n}^{(i)}$,
then we obtain the expression 
\begin{equation}
|\varPhi_{i}\rangle=\sum_{n}\mathsf{C}_{n}^{(i)}|\mathsf{a}_{n},\,\mathsf{b}_{n},\,\mathsf{c}_{n}\rangle
\end{equation}
and complete the proof. $\hfill\blacksquare$

\bibliographystyle{apsrev4-1}
\bibliography{Refs}

\end{document}